\documentclass[a4paper,aps,pre,twocolumn,groupedaddress,showkeys,showpacs,floats,floatats,floatfix,longbibliography]{revtex4-1}
\usepackage[utf8]{inputenc}
\usepackage[english]{babel}
\usepackage{graphicx}
\usepackage{dcolumn} 
\usepackage{bm}
\usepackage{latexsym}
\usepackage{natbib}
\usepackage{mathrsfs}
\usepackage{amssymb}
\usepackage{amsmath, mathtools}
\usepackage{amscd}
\usepackage{color}
\usepackage{xcolor}
\usepackage{pifont}
\usepackage{pstricks,pst-node,pst-text,pst-3d}
\usepackage{verbatim}
\usepackage{soul}
\usepackage[T1]{fontenc}
\usepackage{multirow}
\usepackage{hyperref}
\hypersetup{
  colorlinks   = true, 
  urlcolor     = black, 
  linkcolor    = red, 
  citecolor    = blue 
}
\definecolor{magenta}{rgb}{1.0, 0.0, 1.0}
\definecolor{orange}{rgb}{0.98, 0.6, 0.01}
\definecolor{brown}{rgb}{0.59, 0.29, 0.0}
\definecolor{aquamarine}{rgb}{0.5, 1.0, 0.83}
\definecolor{blue-violet}{rgb}{0.54, 0.17, 0.89}
\definecolor{verde}{rgb}{0.04, 0.6, 0.02}
\definecolor{cinza}{rgb}{0.6, 0.6, 0.6}

\newrgbcolor{Red}{1.0 0.0 1.0}
\begin{document}
\title{Records and occupation time statistics for area-preserving maps$^{\bigstar}$}
\author{Roberto Artuso$^{1,2}$}
\email{roberto.artuso@uninsubria.it}
\author{Tulio M. de Oliveira$^3$}

\author{Cesar Manchein$^3$}
\email{cesar.manchein@udesc.br}
\affiliation{$^1$Dipartimento di Scienza e Alta Tecnologia and Center for Nonlinear and Complex Systems, Via Valleggio 11, 22100 Como, Italy;}
\affiliation{$^2$I.N.F.N, Sezione di Milano, Via Celoria 16, 20133 Milano, Italy;}
\affiliation{$^3$Departamento de F\'\i sica, Universidade do Estado de Santa Catarina, 89219-710 Joinville, SC, Brazil \\
$^{\bigstar}$To Giulio Casati, celebrating his birthday and his achievements.}
\date{\today}
%
\begin{abstract}
  A relevant problem in dynamics is to characterize how
  deterministic systems may exhibit features typically associated to
  stochastic processes. A widely studied example is the study of
  (normal or anomalous) transport properties for deterministic systems
  on a non-compact phase space. We consider here two examples of
  area-preserving maps: the Chirikov-Taylor standard map and the
  Casati-Prosen triangle map, and we investigate transport properties,
  records' statistics and occupation time statistics. While the
  standard map, when a chaotic sea is present, always reproduces
  results expected for simple random walks, the triangle map -whose
  analysis still displays many elusive points- behaves in a wildly
  different way, some of the features being compatible with a
  transient (non conservative) nature of the dynamics.
\end{abstract}

%
\keywords{Area-preserving maps, record statistics, infinite ergodicity.}
\maketitle

\section{Introduction}
One of the most remarkable advances in modern dynamics lies in the
recognition that deterministic systems may exhibit statistical
properties typical of purely stochastic processes: for instance such
systems may display diffusion properties similar to random walks
\cite{LL,Ott,das,RR}. Area-preserving maps (see for instance
\cite{LL}) represent a prominent example of Hamiltonian systems where
subtle features of dynamics, as integrability vs chaotic properties,
may be studied. In this context one of  the most outstanding example
is represented by the Chirikov-Taylor standard map (SM) (see
\cite{st-s,LL} and references therein): we also mention the
fundamental role of such a map in the development of quantum chaos,
unveiling features like quantum dynamical localization
\cite{qc}. Though the SM has been extensively explored by numerical
simulations,  very few rigorous results have been proven (see, for
instance, the introduction in \cite{StL}): however it is generally
believed that for large nonlinearity parameter this map typically
exhibits good stochastic properties, and sensitive dependence upon
initial conditions. Here a remark is due: such a map can be studied
either on a 2-torus or on an (unbounded) cylinder: the latter
representation is naturally adopted when transport properties are
concerned, and analogies with random walks are taken into account
\cite{LL,boris,trans,das}. While particular nonlinear parameters in
the standard map can be tuned to generate strong anomalous diffusion
\cite{vulp}, here we will only deal with the case in which diffusion
is normal. Our findings will be confronted with those obtained for
another area-preserving map, characterized by the lack of exponential
instability: the so called Casati-Prosen Triangle Map (TM) \cite{tm},
introduced by considering, in an appropriate limit, the Birkhoff
dynamics of a triangular billiard: apart from its intrinsic interest,
such a map is an ideal benchmark to test whether stochasticity
properties, exhibited by strongly chaotic systems, are showcased also
by systems lacking any exponential instability. It also turns out that
many features about the TM are still debated, starting from basic
properties like ergodicity and mixing (see for instance
\cite{Mir,Mir2}). 

More precisely we will compare different indicators for both map on
the cylinder:  
though in principle further complications are added when one considers
a non-compact phase space \cite{Aar,Zwe}, this is the appropriate
scenario to discuss transport properties and record statistics, and to
check whether tools from infinite ergodic theory may enrich our
understanding of such systems. 

Our main findings are that the SM, in its typical chaotic regime,
displays all stochastic properties of a purely stochastic system,
while -as expected- results are far more complicated for the TM, even
if we believe that some new insight is provided by our analysis, in
particular as regards persistence behaviour, occupation time
statistics and the relationship between transport properties and
record statistics.

{
The paper is organized as follows. In Sec.~\ref{models}, the
Chirikov-Taylor standard map~\eqref{eq:sm} and the triangle 
map~\eqref{eq:tm} -our basic models- are presented and we also mention
the main properties we analyze. Section~\ref{numerics} is
dedicated to discuss transport properties, records' statistics and occupation time statistics.  We end
with a discussions in Sec.~\ref{conc}.
 } 

\section{The basic setting}
\label{models}
\begin{figure}[h]
   \centering
   \includegraphics[width=7.cm]{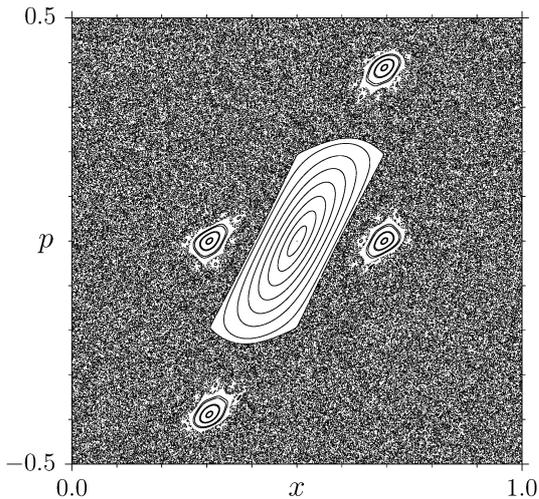}
   \caption{Phase-space portrait for the standard map~\eqref{eq:sm} on
     the 2-torus, for $K=2.6$. Here 40 uniformly distributed initial
     conditions were used for $x$, while maintaining $p_0=0$ fixed: each
     initial condition is iterated $10^4$ times. \label{fig1}}
\end{figure}

We recall the definition of the SM
\begin{equation}
  \begin{array}{ll}
    p_{n+1} = p_{n} + \dfrac{K}{2 \pi}\sin~(2 \pi x_{n}),\\
    x_{n+1} = x_n + p_{n+1} \qquad \mbox{mod}~1;
  \end{array}
  \label{eq:sm}
\end{equation}
$K$ being the nonlinear parameter: when $K$ is sufficiently big no KAM invariant circles bound the motion and one can study moments of the diffusing variable $p \in \mathbb{R}$:
\begin{equation}
\langle \left| p_n -p_0 \right|^q \rangle \sim n^{q \nu(q)}.
\label{eq:Msp}
\end{equation}
The typical behaviour observed for the second moment in simulations is
normal diffusion $\nu(2)=1/2$ \cite{RW, DMP}, while, for certain
parameter values, the existence of stable running orbits (accelerator
modes) induces superdiffusion, $\nu(2)>1/2$) \cite{ishi,benk,zas}. We point
out that a finer description of anomalous transport is obtained by
considering the full spectrum $\nu(q)$: if $\nu(q)=\alpha \cdot q$,
for some $\alpha \neq 1/2$ one speaks about weak anomalous diffusion
whereas the case of a nontrivial $\nu(q)$ is dubbed strong anomalous
diffusion \cite{vulp}. As far as the SM is concerned we will consider
the case where transport in the stochastic sea is normal (even if the
phase space exhibits a mixture of chaotic and elliptic components (see
Figure~\ref{fig1}). 
\unskip

\begin{figure}[h]
    \centering
    \includegraphics*[width=6.8cm]{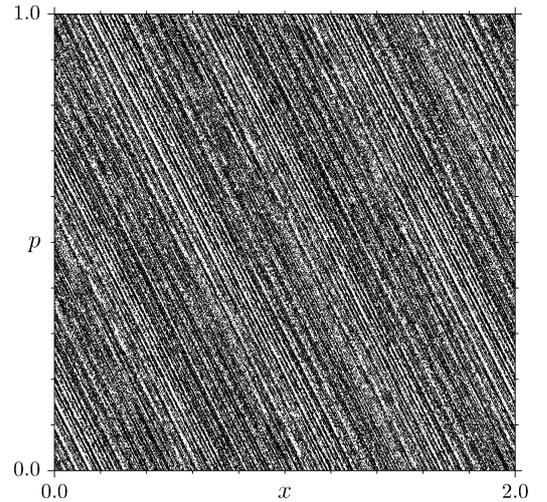}
    \caption{Phase-space dynamics for the triangle map~\eqref{eq:tm}, for 
      $\mu = \frac{1 +\sqrt{5}}{2}$ (golden mean). Here 100 randomly
      distributed initial conditions were used for $x$ and $p$: each
      initial condition is iterated $5 \times 10^4$ times. Notice the
      typical filament structure in the phase space \cite{art1,art2}.}   
    \label{fig2}
  \end{figure}

On the other side the TM is defined (on the cylinder) as:
\begin{equation}
  \begin{array}{ll}
    p_{n+1} = p_n + 2(x_n - \downharpoonright x_n\upharpoonleft - \mu
    (-1)^{\downharpoonright x_n\upharpoonleft }),\\ 
    x_{n+1} = x_n - 2 p_{n+1} \qquad \mbox{mod}~2,
  \end{array}
  \label{eq:tm}
  \end{equation}
  where $\downharpoonright \dots \upharpoonleft$ denotes the nearest
  integer. It was introduced in \cite{tm} (see also \cite{KaHe}) as a
  limit case for the Birkhoff map of irrational triangular billiards:
  systems lacking exponential instability, whose ergodic properties
  are subtly related to irrationality properties of the angles
  \cite{caspro1,art1,art2,PrZn}: we remark that polygonal billiards
  represent both a hard mathematical challenge
  \cite{gut1,gut2,gut3,gut4}, and a natural benchmark when trying to
  assess which microscopic dynamical features lead to macroscopic
  transport laws \cite{Dan,Lam,Sand} ( see also
  \cite{cecc1,cecc2}): in this respect it is worth mentioning that
  anomalous transport has been associated to scaling exponents of the
  spectral measure \cite{agr}, and that generalized triangle maps have
  been investigated recently, both as connected to dynamical
  localization \cite{Kar}, and with respect to slow
  diffusion  \cite{G22}. A typical phase portrait (on the torus) of
  the TM is shown in Figure~\ref{fig2}. 
  \unskip
  Before mentioning the numerical experiments we performed, a crucial
observation is in order. When looking at transport properties (and
records statistics), considering maps on the cylinder is quite
natural, while from the ergodic point of view this perspective is somehow
delicate, since no renormalizable invariant density exists
\cite{Aar,Zwe}, and the appropriate setting is infinite ergodic
theory. When polygonal channels are considered, even establishing
recurrent properties of the dynamics is a demanding task
\cite{Gut-rec}. 

The first set of properties we investigated is more conventional, and
a few results -as we will mention in the next section- have already
been considered, especially as far as the SM is considered. We will look at transport properties, in particular through
the first and the second moment of the diffusing variable We will also
consider records' statistics, which recently has turned very popular
(see \cite{Maj1,Maj2} and references therein). Then we will study
statistical properties like persistence probability and (generalized)
arcsine law \cite{Feller1,Feller2}: while motion in the stochastic sea
for the SM will exhibit typical properties of a simple stochastic
process like a random walk, our findings are quite different in the
case of the TM. 
\section{Results}
\label{numerics}
We start by considering properties associated to the spreading of
trajectories over the phase space, then we will consider occupation
time statistics. 
\subsection{Diffusion}
This is a warm-up exercise, since transport properties have been
studied both for the SM \cite{RW,DMP,LL} and for the TM
\cite{PrZn}. We observe normal transport for the case of the SM (see
panels (c) and (d) in Figure~\ref{rmSM}), while for the TM are results
indicate a superdiffusion, with 
\begin{equation}
  \langle (p_n-p_0)^2 \rangle \sim n^{1.83},
  \label{ADTM}
\end{equation}
in agreement with \cite{PrZn}. We remark that by looking at the
power-law exponents of the first two moments, we have that possibly
anomalous diffusion is weak \cite{vulp}, namely if we consider the
full spectrum of moments' asymptotics: 
\begin{equation}
\label{m-spec}
\langle \left| p_n - p_0 \right|^q \rangle \sim n^{\phi(q)},
\end{equation}
we have a single scaling, in the sense that
\begin{equation}
\label{salpha}
\phi(q)=\alpha \cdot q;
\end{equation}
where normal diffusion is recovered when $\alpha =1/2$.
This is reasonable since weak anomalous diffusion has been observed in 
polygonal billiards \cite{ArtReb}.  
\subsection{Average number of records}
The statistics of records is very popular in the analysis of
correlated and uncorrelated stochastic time sequences
\cite{Maj1,Maj2}: since this subject has not been explores thoroughly
in the deterministic setting (with the remarkable exception of
\cite{Laks1,Laks2}), we briefly review the basic concepts. 
\begin{figure}[!htb]
 \centering
 \includegraphics*[width=0.99\linewidth]{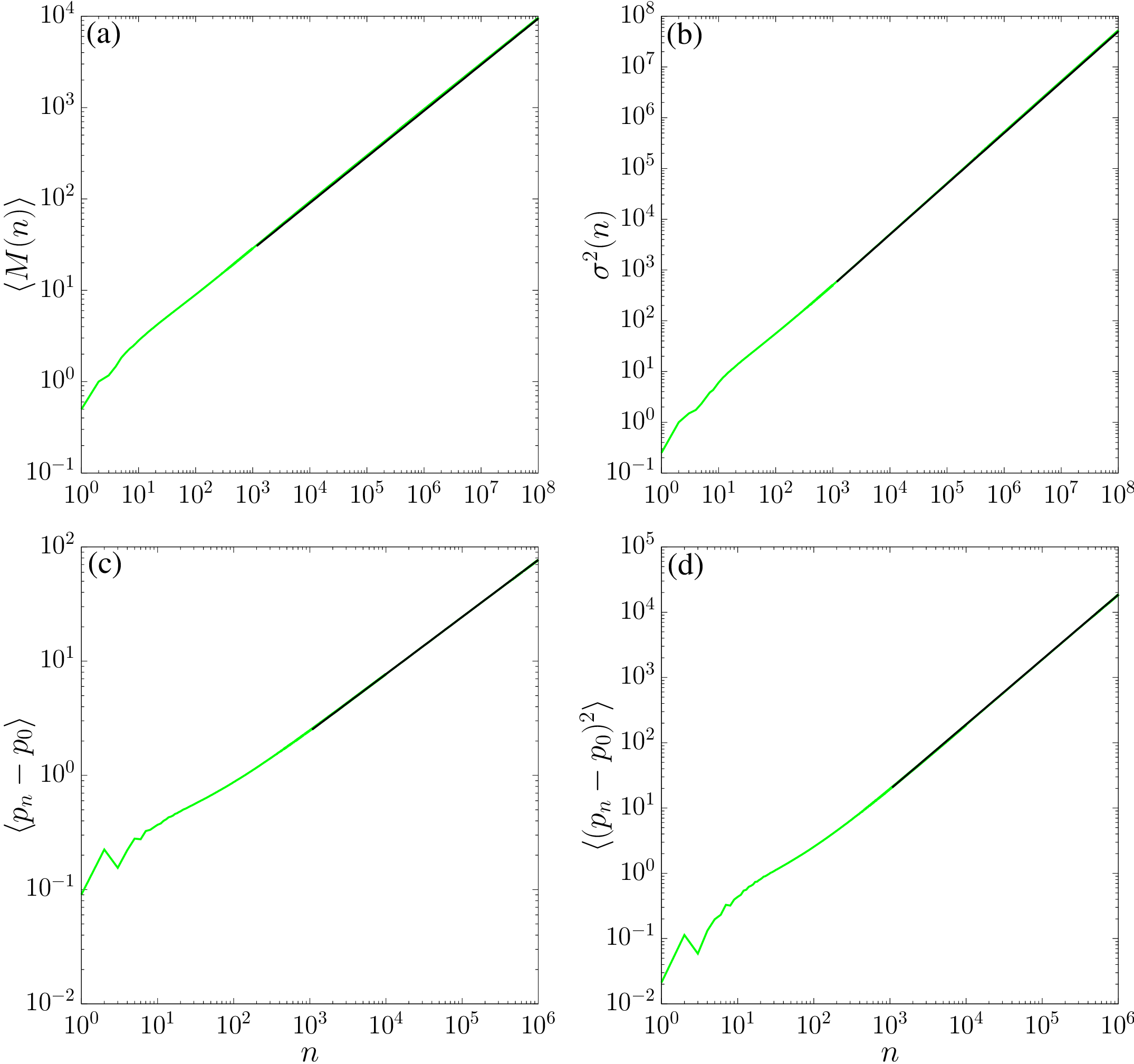}
 \caption{(a) Average number of records, (b) variance, (c) first, and
   (b) second moments of variable $p$ for $K=2.6$ in the standard
   map~\eqref{eq:sm}, as a function of time. These quantities were computed
   for $10^6$ initial conditions for $x_0$, arbitrarily chosen in the
   chaotic sea along the line $p_0=0$. Black-continuous
   lines correspond to power-law asymptotics fit $F(n) = an^\gamma$:
   the fitting parameters are, for (a) $a=0.86(0)$,
   $\gamma=0.50(9)$, for (b) $a = 0.50(7)$, $\gamma=1.00(3)$, for (c)
   $a=0.77(7)$, $\gamma=0.50(1)$ and
   for (d) $a=0.02(1)$, $\gamma = 0.99(1)$. }   
 \label{rmSM}
\end{figure}
First of all let us recall the (straightforward) definition of a
record: given a sequence of real data $x_0,\,x_1, \, \dots , x_k,\,
\dots$ the element $x_m$ is a record if 
\begin{equation}
x_m > x_j \qquad j=0,\,1,\, \dots m-1,
\label{rec-def}
\end{equation}
(we consider $x_0$ to be the first record). To the sequence of data
points we associate the binary string $\sigma_0,\,\sigma_1,\, \dots ,
\sigma_k,\, \dots$, where 
\begin{equation}
\sigma_l=
\left\{
\begin{array}{l}
1 \qquad \mbox{if $x_l$ is a record} \\
0 \qquad \mbox{otherwise}
\end{array}
\right.
\label{sigmarec}
\end{equation}
The number of records up to time $N$ is then
\begin{equation}
M_N=\sum _{j=0}^N\, \sigma_j.
\label{srec}
\end{equation}

\begin{figure}[!htb]
 \centering
 \includegraphics*[width=0.99\linewidth]{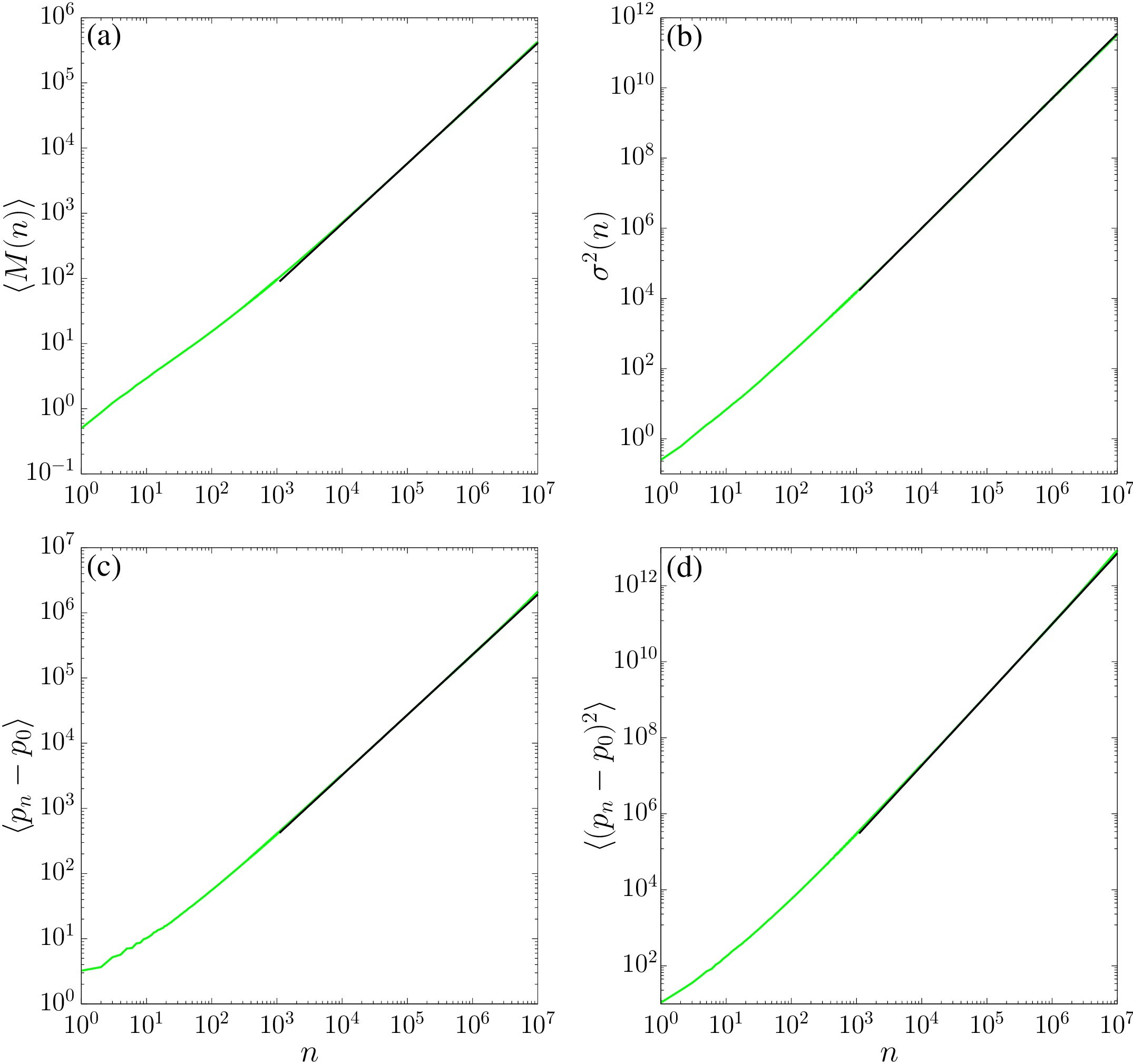}
 \caption{(a) Average number of records, (b) variance, (c) first, and 
   (b) second moments of variable $p$ for the golden mean 
   $\mu=\frac{1+\sqrt{5}}{2}$ in the triangular map~\eqref{eq:tm} as a 
   function of time. These quantities were computed for $10^6$ initial 
   conditions for $x_0$, arbitrarily chosen in phase space along the
   line $p_0=0$. Black-continuous lines correspond to
   power-law asymptotics $F(n)=an^\gamma$: the fitting parameters are,
   for (a)  $a=0.13(9)$, $\gamma = 0.92(4)$, for (b) $a=0.04(0)$,
     $\gamma=1.84(9)$, for (c) $a=0.65(4)$, $\gamma=0.92(4)$ and 
   for (c) $a=0.67(6)$, $\gamma =1.86(0)$ in  (d).}     
 \label{rmTM}
\end{figure}

The properties of the average number of records, $\langle M_N
\rangle$, and the corresponding variance 
\begin{equation}
\label{varM}
Var(M_N)=\langle M_N^2 \rangle - \langle M_N \rangle^2
\end{equation}
are important tools to access the nature of the data sequence: as a matter of fact if the different $x_j$ are independent, identically distributed random variables, then, for large $N$ we have \cite{Wer,Nev}:
\begin{equation}
\label{Miid}
\langle M_N\rangle =\ln N +\gamma_E +{\cal O}(N^{-1}),
\end{equation}
where  $\gamma_E=0.5772\dots$ is the Euler-Mascheroni constant, and 
\begin{equation}
\label{Viid}
Var(M_N)=\sigma^2(N)=\ln N +\gamma_E-\frac{\pi^2}{6}+{\cal O}(N^{-1}).
\end{equation}
We remark that both quantities are independent of the common distribution of the random variables: this universality is an important feature of record statistics in different contexts.\\
Results are quite different for a correlated sequence, as when $x_j$ denotes the position of a random walker at time $j$:
\begin{equation}
\label{xrw}
x_{j+1}=x_j+\xi_{j+1},
\end{equation}
where the jumps are taken from a common distribution $\wp(\xi)$.
In this case the behaviour is \cite{Maj1,Maj2}:
\begin{equation}
\label{Mrw}
\langle M_N \rangle \approx \frac{2}{\sqrt{\pi}} \sqrt{N},
\end{equation}
and
\begin{equation}
\label{M2rw}
Var(M_N)\approx 2 \left( 1- \frac 2{\pi}\right) N,
\end{equation}
so that the standard deviation is of the same order of magnitude as
the average. Again this is a {\it universal} result,
independent of the particular jump distribution $\wp(\xi)$, as long as
the distribution is continuous and symmetric. The crucial ingredient
of the proof is that the process renews as soon as a new record is
achieved and the appearance of the new record is related to the
survival probability for the process, which is universal in view of
Sparre-Andersen theorem \cite{Feller2,SpAn1,SpAn2} (see also
\cite{geor}).  

\begin{figure}[!htb]
  \centering
  \includegraphics*[width=0.8\linewidth]{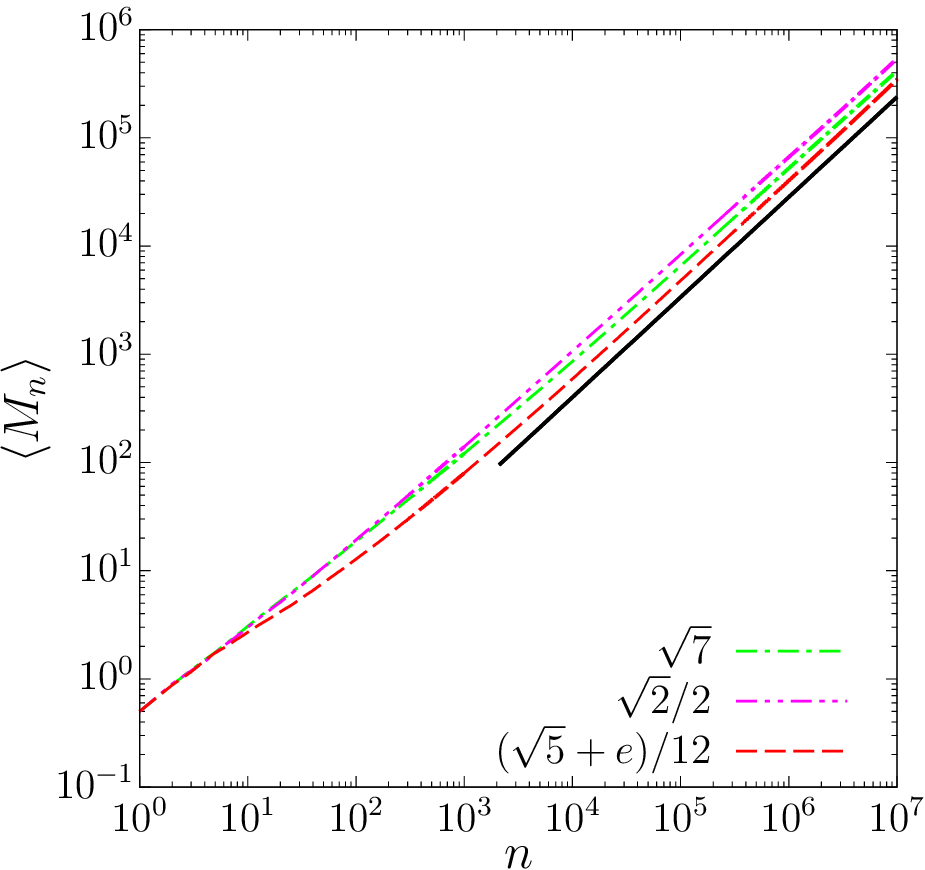}
  \caption{(a) Average number of records for three
    additional parameters $\mu$ in the TM. 
      Black-continuous line correspond to the power-law  asymptotic
    fitting function $F(n)=an^\gamma$, with $\gamma = 0.92(4)$.}   
  \label{recmany}
\end{figure}

Numerical results on records statistics are reported in 
  Figures~\ref{rmSM}, \ref{rmTM}, panels (a) and (b): for the SM
our results are consistent with early investigations \cite{Laks1,Laks2},
and with the asymptotic behaviour of a random walk, while for the TM
we observe anomalous scaling w.r.t. (\ref{Mrw},\ref{M2rw}): the
behaviour is related to transport properties, in the sense that data
are consistent with the growths: 
\begin{equation}
\label{mom-rec}
\langle M_N \rangle \sim N^{\phi(1)}, \qquad Var(M_N) \sim N^{\phi(2)}.
\end{equation}
A similar behaviour was observed in \cite{Laks1,Laks2}, for the SM in
the presence of accelerator modes. We remark that, though in the
following we will fix our attention of a particular parameter value
for the TM, we checked that reported experiments do not depend on the
particular parameter choice, as exemplified in Figure~\ref{recmany},
where the growth of the averaged number of records is reported for
different parameters of the TM.  
\\
While a general, quantitative relationship (if any) between transport
exponents and statistical properties of records has not been fully
developed, to the best of our knowledge, it is possible, in some
cases, to connect $\phi(1)$ to the expected maximum of the walk
\cite{Maj3,Maj4}, that, for random walk with unit jumps, coincides
with the number of records. On the other side we mention that
non-homogeneous random walks offer examples where such relationship
does not hold \cite{Gil,MSer, age,Sin,Last}. 

\subsection{Occupation time statistics}

When we consider the evolution on the cylinder, both for the SM and
the TM, we are in the presence of infinitely ergodic systems
\cite{Aar,Zwe}, since, while Lebesgue measure is preserved, due to
area conservation, the (constant) phase space is unbounded, so the
invariant density cannot be normalized. This has a series of
remarkable consequences, which originally have been considered in the
context of stochastic processes, and then explored in the 
deterministic evolution framework. 

One of the most striking property that has been investigated is the
generalized arcsine law (see \cite{Feller1} for the standard
formulation for stochastic processes): we briefly recall the main
result that lies at the basis of our analysis, namely Lamperti's
theorem \cite{Lamp}. The original formulation involves discrete
stochastic processes, for which the infinite set of possible states
can be separated into two sets $A$ and $B$ separated by a single site
$x_0$, such that a transition from one set to the other can only be
achieved by passing through $x_0$, which can be taken as the starting
site, and is supposed to be recurrent (namely the probability of
returning to it is 1). For instance we can think of one dimensional
random walk on an integer lattice, with $x_0=0$ and $A$ ($B$) consists
of strictly positive (negative) lattice sites. We are interested in
the limiting distribution of $N(n)/n$, the fraction of time spent in
the positive semi-axis up to time $n$. The theorem states that such a
distribution exists in the $n \to \infty$ limit, and it is
characterized by two parameters $\alpha$ and $\eta$. $\eta$ is related
to symmetry properties of the process, being the expectation value of
the fraction of time spent in $\mathbb{R}_+$: 
\begin{equation}
\label{etaL}
\eta =\lim_{n \to \infty}\mathbb{E}\left(\frac{N(n)}{n}\right ) :
\end{equation}
for a symmetric process $\eta=1/2$, and from now on we will only
consider such a case. 

\begin{widetext}
  $\quad$
\begin{figure}[!htb]
 \centering
 \includegraphics*[width=0.99\linewidth]{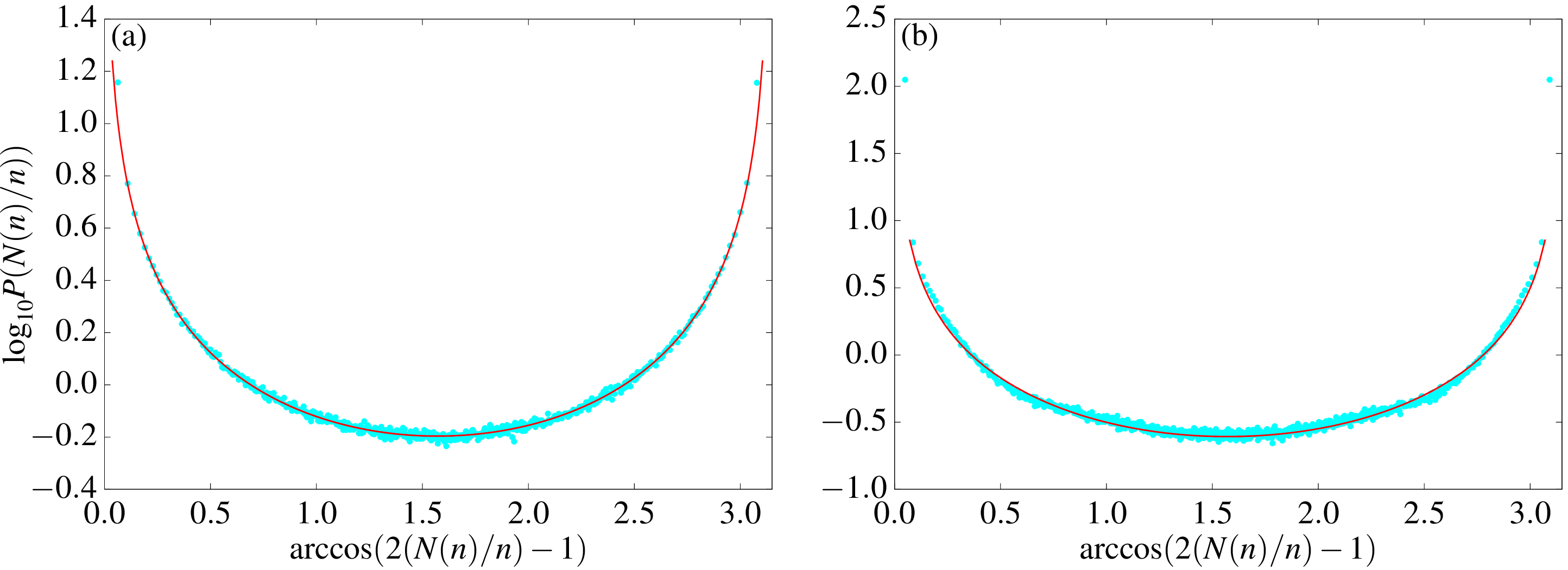}
 \caption{Distribution of the fraction of time spent in the positive
   axis for the momentum $p$ in the standard ~\eqref{eq:sm} (a) and
   triangle ~\eqref{eq:tm} (b) maps, in semi-logarithmic scale. To
   enhance readability of the border values, the transformation
   $x\rightarrow \arccos(2x-1)$ on the horizontal axis. The (light blue) 
   points represent the simulation results, the (red) line the
   Lamperti distribution \eqref{LDist}. Data are obtained by computing $10^6$ initial
   conditions iterated $10^6$ times for the standard map and $10^6$
   initial conditions iterated $10^8$ times for the triangle map. The
   fitting parameters are $\alpha = 0.49(9)$ for (a) and 
   $\alpha = 0.42(0)$ for (b). In the case of the TM,
     data suggest a superposition of a
     (rescaled) Lamperti distribution and two Dirac's $\delta$
     centered of $x=0$ and $x=1$ (see text).}
 \label{ldSM}
\end{figure}
\end{widetext}

The other parameter $\alpha$ is instead connected to the behaviour of
the generating function of first return probabilities to the starting
site: it can be shown \cite{roap} that it can be related to the
probability $P_n$ of being at the starting site after $n$ steps in the
following way: 
\begin{equation}
\label{Lret}
P_n \sim \frac{H(n)}{n^{1-\alpha}},
\end{equation}
where $H(n)$ is a slowly varying function, namely
\begin{equation}
\label{slow}
\lim_{n\to \infty} \frac{H(yn)}{n}=1.
\end{equation}
Under such conditions the density of $\varphi=N(n)/n$ in the infinite
time limit is given by Lamperti distribution: 
\begin{equation}
\label{LDist}
{\cal G}_{\alpha}(\varphi)= \frac{\sin(\pi
  \alpha)}{\pi}\frac{\varphi^{1-\alpha}(1-\varphi)^{1-\alpha}}{\varphi^{2\alpha}+2
  \varphi^{\alpha}(1-\varphi)^{\alpha}\cos(\pi \alpha)
  +(1-\varphi)^{2\alpha}}, 
\end{equation}
that reproduces the usual arcsine law 
\begin{equation}
\label{arc-sin}
\mathbb{P}\left((N_n/n)\leq \xi \right)=\frac{2}{\pi}\arcsin\left( \sqrt{\xi}\right)
\end{equation}
when $\alpha=1/2$, in the universality class of Sparre-Andersen
theorem. Deviations from standard arcsine law have been reported
for a number of cases, in the framework of deterministic dynamics
\cite{BeBa1,BMS,Thal1,ZwL,hz,Thal2,Ak,SK}, mainly in the context of
intermittent maps. 
Numerical experiments for the SM confirm the validity of the arcsine
law, $\alpha=1/2$, see panel (a) in Figure~\ref{ldSM}: to our
knowledge this is the first time such an indicator has been considered
in the analysis of area preserving maps.\\ 
The results, as expected, are quite different for the TM, and they
suggest novel features exhibited by this map. In particular (see panel (b) in Figure~\ref{ldSM}) numerical results
are well fitted by a Lamperti distribution (with $\alpha \approx 0.42$), thus different from an ordinary random
walk), except for the endpoints, that present enhanced
peaks. Intuitively such an additional contribution might be due to a
fraction of orbits never returning to the origin: this would
correspond, in stochastic language, to a transient random walk (we
recall that, according to P\'olya's theorem \cite{Hugh} a simple
symmetric random walk is recurrent -so the return to the starting site
is sure- in one and two dimensions, and transient in higher
dimensions). Such a possibility is indeed not excluded for infinite
polygonal channels \cite{Gut-rec}. \\Our last set of simulations
concerns the survival probability \cite{BMS}: 
\begin{equation}
\label{Psurv}
P_{cum}(n)=\mathrm{prob}\left( p_n\geq 0 \dots p_1 \geq 0 | p_0=0 \right).
\end{equation}

\begin{widetext}
  $\quad$
\begin{figure}[!htb]
  \centering
  \includegraphics*[width=0.75\linewidth]{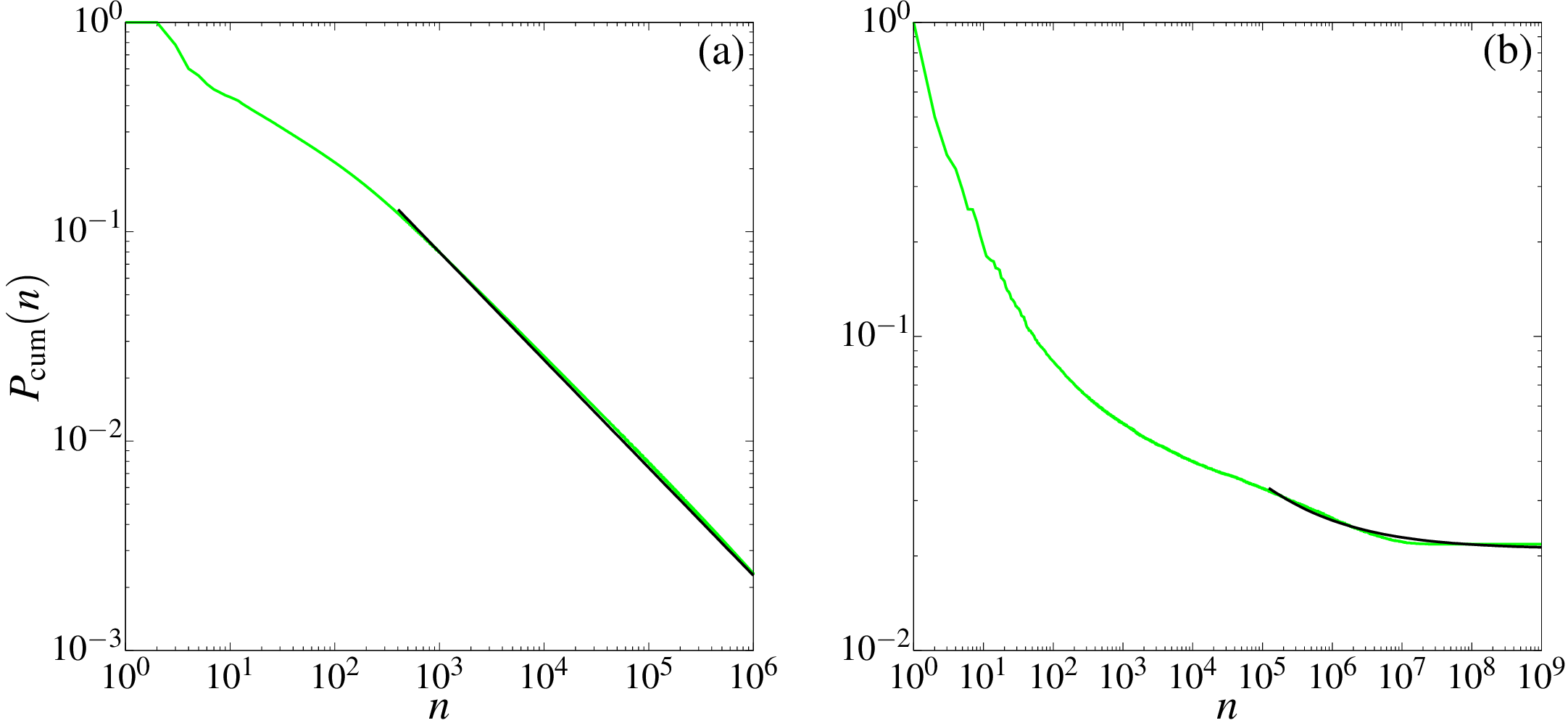}
  \caption{Cumulative distribution function for the
      survival times obtained for the variable $p$ for (a) the standard 
      map~\eqref{eq:sm},  and (b) the triangle map~\eqref{eq:tm}, in
      logarithmic scale. Data are obtained by simulating $10^6$ and $10^5$
      initial conditions, respectively. Continuous-black lines correspond
      to power-law asymptotic functions $F(n)=a+bn^{-\alpha}$: the
      fitting  parameters are $a=0,~b=2.80(0)$, and $\alpha = 0.51(5)$
      in (a) and $a=0.021(0),~b=1.62(6)$, and $\alpha = 0.42(0)$ in (b).} 
  \label{survPlot}
\end{figure}
\end{widetext}

When considering recurrent random walks, the asymptotic behaviour of
the survival probability is again ruled by Lamperti exponent
\cite{Lamp,roap} (see also \cite{Eli}): 
\begin{equation}
\label{survL}
P_{cum}(n)\sim n^{-\alpha}.
\end{equation}
Once again SM simulations (see panel (a) in Figure ~\ref{survPlot})
agree with expected behaviour for simple random walks ($\alpha=1/2$),
while the situation is completely different for the
  TM, where the survival probability seems to tend to a finite limit
  for large $n$, see panel (b) in Figure~\ref{survPlot}. This is
  coherent with the transient nature of the TM, which we conjectured
  in the analysis of generalized arcsine law.

\section{Discussion}
\label{conc}
We have performed a set of extensive numerical experiments on two
paradigmatic area-preserving maps, the SM and the TM, focusing in the
case where such maps are considered on a cylinder, namely a non
compact phase space. Firstly we reproduced known results about normal
diffusion for typical (chaotic) parameters of the SM, and
superdiffusion for the TM. Then we explored records' statistics:
numerical simulations again confirm that the SM behave like a simple
random walk, while anomalous growth is exhibited by
the TM. The most interesting results arise in  the analysis of
occupation times, like generalized arcsine law and survival
probability. While once again normal stochastic properties are
displayed by the SM, the TM presents more surprising results, which we
conjecture are possibly connected to lack of conservativity
\cite{Gut-rec} (or transient behaviour, in the language of random
walks). This feature, which we think is worth of further
investigations, might suggest new stochastic modeling
of the TM (see \cite{G22}). 

\vspace{5mm}

\section*{Authors'~contributions}
All authors have contributed substantially to the work. All authors have read and agreed to the published version of the manuscript.

\section*{Acknowledgements}
R.A. acknowledges partial support from PRIN Research Project No. 2017S35EHN ``Regular and stochastic behavior in dynamical systems'' 
of the Italian Ministry of Education, University and Research (MIUR). R.A. acknowledges an association to the GNFM group of
INDAM. R.A thanks Gaia Pozzoli for discussions. C.M. acknowledges the National Council for Scientific and
Technological Development - CNPq (Brazilian agency) for partial financial support (Grant Number 310228/2020-4). T.M.O. acknowledges 
the Coordena\c{c}\~ao de Aperfei\c{c}oamento de Pessoal de N\'ivel Superior - CAPES (Brazilian agency ) - Finance Code 001, 
for partial financial support. Additionally, T.M.O. and C.M. also acknowledges the Funda\c{c}\~ao de Amparo \`a Pesquisa e 
Inova\c{c}\~ao do Estado de Santa Catarina - FAPESC (Brazilian agency) for partial financial support.



\begin{thebibliography}{999}
\bibitem{LL} Lichtenberg, A.J.; Lieberman, M.A. \textit{Regular and chaotic dynamics}; Springer: Berlin, Germany, 1992.
\bibitem{Ott} Ott, E., \textit{Chaos in dynamical systems}; CUP: Cambridge, U.K. 2002.
\bibitem{das} Cvitanovi\'c, P.; Artuso, R.; Mainieri, R.; Tanner, G.; Vattay, G.,
     \textit{Chaos: Classical and Quantum},
    {\tt ChaosBook.org} ; Niels Bohr Institute, Copenhagen 2020.
\bibitem{RR} Artuso, R.; Burioni, R. Anomalous diffusion: deterministic and stochastic perspectives. In {\em Large deviations in physics}; Vulpiani A. et al., Eds.; Springer-Verlag: Berlin Heidelberg, Germany, 2014; pp. 263--293.
\bibitem{st-s} Chirikov, B.V.; Shepelyansky, D.L. Chirikov standard map. {\em Scholarpedia} {\bf 2008}, {\em 3},  3550.
\bibitem{qc} Casati, G.; Chirikov, B.V., (Eds.) {\em Quantum chaos}; OUP: Oxford, U.K., 1995.
\bibitem{StL} Bloor, K.; Luzzatto, S. Some remarks on the geometry of the standard map. {\em Int.J.Bifurcat.Chaos} {\bf 2009}, {\em 19}, 2213--2232.
\bibitem{boris} Chirikov, B.V. A universal instability of many dimensional oscillator systems. {\em Phys.Rep.} {\bf 1979}, {\em 52}, 263--379.
\bibitem{trans} MacKay, R.S.; Meiss,J.D.; Percival, I.C. Stochasticity and transport in hamiltonian systems. {\em Phys.Rev.Lett.} {\bf 1984}, {\em 52}, 697--700.
\bibitem{vulp} Castiglione, P.; Mazzino, A.; Muratore-Ginanneschi, P.; Vulpiani, A, On strong anomalous diffusion. {\em Physica D} {\bf 1999}, {\em 134}, 75--93.
\bibitem{tm} Casati, G.; Prosen, T.. Triangle map: a model of quantum chaos. {\em Phys.Rev.Lett.} {\bf 2000}, {\em 85}, 4261--4264.
\bibitem{Mir} Horvat, M.; Degli Esposti, M.; Isola, S.; Prosen, T.; Bunimovich, L. On ergodic and mixing properties of the triangle map. {\em Physica D} {\bf 2009}, {\em 238}, 395--415.
\bibitem{Mir2} Degli Esposti, M.; Galatolo, S. Recurrence near given sets and the complexity  of the Casati-Prosen map. {\em Chaos, Solitons \& Fractals} {\bf 2005}, {\em 23}, 1275--1284.
\bibitem{Aar} Aaronson, J. \textit{An introduction to infinite ergodic theory}; AMS: Providence, U.S.A. , 1997.
\bibitem{Zwe} Zweim\"uller, R. Surrey notes on infinite ergodic theory. Available online\\
https://mat.univie.ac.at/\%7Ezweimueller/PapersAndPreprints.html (accessed on 1 11 2022)
\bibitem{RW} Rechester, A.B.; White, R.B. Calculation of turbulent diffusion for the Chirikov-Taylor model. {\em Phys.Rev.Lett.} {\bf 1980}, {\em 44}, 1586--1589.
\bibitem{DMP} Dana, I.; Murray, N.W.; Percival, I.C. Resonances and diffusion in periodic Hamiltonian maps. {\em Phys.Rev.Lett.} {\bf 1989}, {\em 62}, 233--236.
\bibitem{ishi} Ishizaki, R.; Horita, T.; Kobayashi, T.; Mori, H. Anomalous diffusion due to accelerator modes in the standard map. {\em Progr.Theor.Phys.} {\bf 1991}, {\em 85}, 1013--1022.
\bibitem{benk} Benkadda, S.; Kassibrakis, S.; White, R.B.; Zaslavsky, G.M. Self-similarity and transport in the standard map. {\em Phys.Rev. E} {\bf 1997}, {\em 55}, 4909--4917.
\bibitem{zas} Zaslavsky, G.M.; Edelman, M.; Niyazov, B.A. Self-similarity, renormalization, and phase-space nonuniformity of Hamiltonian chaotic dynamics. {\em Chaos} {\bf 1997}, {\em 7}, 159--181.
\bibitem{KaHe} Kaplan, L.; Heller, E.J. Weak quantum ergodicity. {\em Physica D} {\bf 1998}, {\em 121}, 1--18.
\bibitem{caspro1} Casati, G.; Prosen, T. Mixing properties of triangular billiards. {\em Phys.Rev.Lett.} {\bf 1999}, {\em 83}, 4728--4732.
\bibitem{art1} Artuso, R.; Casati, G.; Guarneri, I. Numerical study on ergodic properties of triangular billiards. {\em Phys.Rev. E} {\bf 1997}, {\em 55}, 6384--6390.
\bibitem{art2} Artuso, R. Correlations and spectra of triangular billiards. {\em Physica D} {\bf 1997}, {\em 109}, 1--10.
\bibitem{PrZn} Prosen, T.; \u{Z}nidari\^{c}, M. Anomalous diffusion and dynamical localization in polygonal billiards. {\em Phys.Rev.Lett.} {\bf 2001}, {\em 87}, 114101
\bibitem{gut1} Gutkin, E. Billiards in polygons. {\em Physica D} {\bf 1986}, {\em 19}, 311--333.
\bibitem{gut2} Gutkin, E. Billiards in polygons: survey of recent results. {\em J.Stat.Phys.} {\bf 1996},  {\em 83}, 7--26.
\bibitem{gut3} Gutkin, E. Billiard dynamics: a survey with the emphasis on open problems. {\em Regular and chaotic dynamics} {\bf 2003}, {\em  8}, 1--13.
\bibitem{gut4} Gutkin, E. Billiard dynamics: an updated survey with the emphasis on open problems. {\em Chaos} {\bf 2012}, {\em 22}, 026116.
\bibitem{Dan} Alonso, D.; Ruiz, A.; De Vega, I. Transport in polygonal billiards. {\em Physica D} {\bf 2004}, {\em 187}, 184--199.
\bibitem{Lam} Jepps, O.G.; Bianca, C.; Rondoni, L. Onset of diffusive behavior in confined transport systems. {\em Chaos} {\bf 2008}, {\em 18}, 013127.
\bibitem{Sand} Sanders, D.P.; Larralde, H. Occurrence of normal and anomalous diffusion in polygonal billiard channels. {\em Phys.Rev. E} {\bf 2006}, {\em 73}, 026205.
\bibitem{cecc1} Cecconi, F.; Del-Castillo-Negrete, D.; Falcioni, M.; Vulpiani, A. The origin of diffusion: the case of non-chaotic systems. {\em Physica D} {\bf 2003}, {\em 180}, 129--139.
\bibitem{cecc2} Cecconi, F.; Cencini, M.; Falcioni, M.; Vulpiani, A. Brownian motion and diffusion: from stochastic processes to chaos and beyond. {\em Chaos} {\bf 2005}, {\em 15}, 026102.
\bibitem{agr} Artuso, R.; Guarneri, I.; Rebuzzini, L. Spectral properties and anomalous transport in a polygonal billiard. {\em Chaos} {\bf 2000}, {\em 10}, 189--194.
\bibitem{Kar} Guarneri, I.; Casati, G.; Karle, V. Classical dynamical localization. {\em Phys.Rev.Lett.} {\bf 2014}, {\em 113}, 174101
\bibitem{G22} Yoshida, K.; Casati, G.; Watanabe, S.; Shudo, A. Sublinear diffusion in the generalized triangle map. {\em Phys.Rev. E} {\bf 2022}, {\em 106}, 014206.
\bibitem{Gut-rec} Conze, J-P.; Gutkin, E. On recurrence and ergodicity for geodesic flows on non-compact periodic polygonal surfaces {\em Ergod.Th.\& Dynam.Sys} {\bf 2012}, {\em 32}, 491--515.
\bibitem{Maj1} Majumdar, M.N. Universal first-passage properties of discrete-time random walks and L\'evy flights on a line: statistics of the global maximum and records. {\em Physica A} {\bf 2010}, {\em 389}, 4299--4316.
\bibitem{Maj2} Godr\`eche, C.; Majumdar, S.N.; Schehr, G. Record statistics of a strongly correlated time series: random walks and L\'evy flights. {\em J.Phys. A} {\bf 2017}, {\em 50}, 333001.
\bibitem{Feller1} Feller, W. \textit{An introduction to probability theory and its applications. Vol. 1}; Wiley: New York, U.S.A., 1968.
\bibitem{Feller2} Feller, W. \textit{An introduction to probability theory and its applications. Vol. 2}; Wiley: New York, U.S.A., 1971.
\bibitem{ArtReb} Rebuzzini, L.; Artuso, R. Higher order statistics in the annulus square billiard: transport and polyspectra. {\em J.Phys. A} {\bf 2011}, {\em 44}, 025101.
\bibitem{Laks1} Srivastava, S.C.L.; Lakshminarayan, A.; Jain, S.R. Record statistics in random vectors and quantum chaos. {\em Europhys.Lett.} {\bf 2013}, {\em 101}, 10003.
\bibitem{Laks2} Srivastava, S.C.L.; Lakshminarayan, A. Records in the classical and quantum standard map. {\em Chaos, Solitons \& Fractals} {\bf 2015}, {\em 74}, 67--78.
\bibitem{Wer} Wergen, G. Records in stochastic processes: theory and applications. {\em J.Phys. A} {\bf 2013}, {\em 46}, 223001.
\bibitem{Nev} Nevzorov, V.B. \textit{Records: mathematical theory}; AMS: Providence, U.S.A., 2004.
\bibitem{SpAn1} Sparre Andersen, E. On the fluctuations of sums or random variables I. {\em Math.Scand.} {\bf 1953}, {\em 1}, 263--285.
\bibitem{SpAn2} Sparre Andersen, E. On the fluctuations of sums or random variables II. {\em Math.Scand.} {\bf 1954}, {\em 2}, 195--233.
\bibitem{geor} Artuso, R.; Cristadoro, G.; Degli Esposti, M.; Knight, G. Sparre-Andersen theorem with spatiotemporal correlations. {\em Phys.Rev. E} {\bf 2014}, {\em 89}, 052111.
\bibitem{Maj3} Comtet, A.; Majumdar, S.N. Precise asymptotics for a random walker's maximum. {\em J.Stat.Mech.} {\bf 2005}, P06013.
\bibitem{Maj4}Mounaix, P.; Majumdar, S.N.; Schehr, G. Asymptotics for the expected maximum of ramndom walks and L\'evy flights with a constant drift. {\em J.Stat.Mech.} {\bf 2018}, P083201.
\bibitem{Gil} Gillis, J. Centrally biased discrete random walk. {\em Q.J. Math.} {\bf 1956}, {\em 7}, 144--152.
\bibitem{MSer} Serva, M. Scaling behavior for random walks with memory of the largest distance from the origin. {\em Phys. Rev. E} {\bf 2013}, {\em 88}, 052141.
\bibitem{age} Radice, M.; Onofri, M.; Artuso, R.; Cristadoro, G. Transport properties and ageing for the averaged L\'evy-Lorentz gas. {\em  J. Phys. A} {\bf 2020},{\em 53}, 025701.
\bibitem{Sin} Singh, P. Extreme value statistics and arcsine laws for heterogeneous diffusion processes. {\em Phys. Rev. E} {\bf 2022}, {\em 105}, 024113.
\bibitem{Last} Artuso, R.; Onofri, M.; Pozzoli, G.; Radice, M. Extreme value statistics of positive recurrent centrally biased random walks. {\em J.Stat.Mech.} {\bf 2022}, P103209.
\bibitem{Lamp} Lamperti, J. An occupation time theorem for a class of stochastic processes. {\em \rm Trans.Amer.Math.Soc} {\bf 1958}, {\em 88}, 380--387.
\bibitem{roap} Radice, M.; Onofri, M.; Artuso, R.; Pozzoli, G. Statistics of occupation times and connection to local properties of nonhomogeneous random walks. {\em Phys.Rev. E} {\bf 2020}, {\em 101}, 042103.
\bibitem{BeBa1} Bel, G.; Barkai, E. Weak ergodicity breaking with deterministic dynamics. {\em Europhys.Lett.} {\bf 2006}, {\em 74}, 16--21.
\bibitem{BMS} Bray, A.J.; Majumdar,  S.N.; Schehr, G. Persistence and first-passage properties in nonequilibrium systems. {\em Adv.Phys.} {\bf 2013}, {\em 62}, 225--361.
\bibitem{Thal1} Thaler, M. The Dynkin-Lamperti arc-sine laws for measure preserving transformations. {\rm Trans.Amer.Math.Soc} {\bf 1998}, {\em 350}, 4593--4607.
\bibitem{ZwL} Zweim\"uller, R. Infinite measure preserving transformations with compact first regeneration. {\em Jour.Anal.Math.} {\bf 2007}, {\em 103}, 93--131.
\bibitem{hz} Huang, J.; Zhao, H. Ultraslow diffusion and weak ergodicity breaking in right triangular billiards. {\em Phys.Rev. E} {\bf 2017}, {\em 95}, 032209.
\bibitem{Thal2} Thaler, M. A limit theorem for sojourns near indifferent fixed points of one dimensional maps. {\em Ergod.Theory \& Dyn.Syst.} {\bf 2002}, {\em 22}, 1289--1312.
\bibitem{Ak} Akimoto, T. Generalized arcsine law and stable law in an infinite measure dynamical system. {\em J.Stat.Phys.} {\bf 2008}, {\em 132}, 171--186.
\bibitem{SK} Singh, P.; Kundu, A. Generalized `arcsine' laws for run-and-tumble particle in one dimension. {\em J.Stat.Mech.} {\bf 2019}, 083205.
\bibitem{Hugh} Hughes, B.D. \textit{Random walks and random environments. Volume I: Random walks}; Clarendon Press: Oxford, U.K., 1995.
\bibitem{Eli} Barkai, E. Residence time statistics for normal and fractional diffusion in a force field. {\em J.Stat.Phys.} {\bf 2006}, {\em 123}, 883--907.
\end{thebibliography}


\end{document}